\documentclass{PoS}

\usepackage{amsmath}
\usepackage{comment}

\title{Phase structure analysis of ${\rm CP}(N-1)$ model using Tensor renormalization group}

\ShortTitle{Phase structure analysis of CP($N-1$) model using Tensor renormalization group}

\author{\speaker{Hikaru Kawauchi}
\\
        Institute for Theoretical Physics, Kanazawa University, Kanazawa 920-1192, Japan\\
        E-mail: \email{kawauchi@hep.s.kanazawa-u.ac.jp}}

\author{Shinji Takeda\thanks{
       This work was supported by Kanazawa University SAKIGAKE Project
and by MEXT as "Exploratory Challenge on Post-K computer" (Frontiers of Basic Science: Challenging the Limits).
}
\\
        Institute for Theoretical Physics, Kanazawa University, Kanazawa 920-1192, Japan\\
        E-mail: \email{takeda@hep.s.kanazawa-u.ac.jp}}

\abstract{The phase structure of the lattice CP($N-1$) model in two dimensions is analyzed by the tensor renormalization group (TRG) method. We focus on the case $N=2$ and compare the numerical result of the TRG method with that of the strong-coupling analysis in the presence of the $\theta$ term and investigate the nature of the phase transition at $\theta=\pi$.
          }

\FullConference{34th annual International Symposium on Lattice Field Theory\\
		24-30 July 2016\\
		University of Southampton, UK}

\begin{document}

\section{Introduction}
The two dimensional CP($N-1$) model is a toy model of QCD since they have a lot in common; for example, they are asymptotic free theory and include the $\theta$ term. In QCD, the difficulty in understanding why the value of $\theta$ is so small is one of the remaining puzzles, which is called the strong CP problem. While solving this problem in QCD is significant, analyzing the CP($N-1$) model in the presence of the $\theta$ term is meaningful too.

\begin{figure}[h]
 \centering
  \includegraphics[width=80mm]{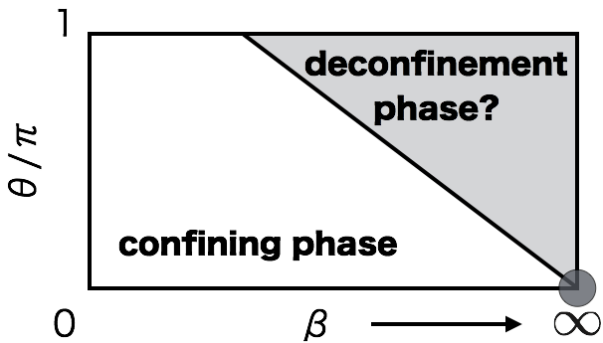}
 \caption{The phase diagram of the CP($N-1$) model proposed by Schierholtz. If the deconfinement phase exists and $\theta=0$ is the only choice of the confining phase in the continuum limit, then the strong CP problem in the CP($N-1$) model would be resolved.}
 \label{schierholz}
\end{figure}

\begin{figure}[h]
 \centering
  \includegraphics[width=90mm]{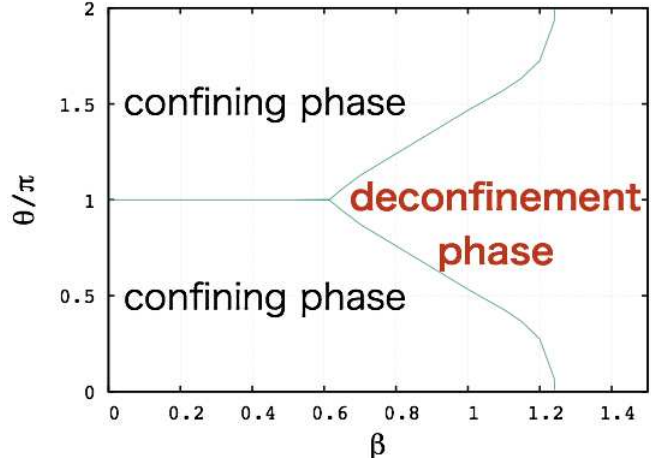}
 \caption{The phase diagram of the CP(1) model by using the strong coupling analysis~\cite{Plefka}. The deconfinement phase appears and supports Schierholz's scenario.}
 \label{strongcoupling}
\end{figure}

In the lattice CP($N-1$) model, an interesting scenario that $\theta$ becomes zero was proposed by Schierholz in Ref.~\cite{Schierholz}. He analyzed the phase diagram of the CP($N-1$) model in the $\beta$-$\theta$ plane as shown in Fig.~\ref{schierholz}. In the phase diagram, a phase transition from confining phase to deconfinement phase was found and he suggested that if $\theta=0$ is the only point in the confining phase at which the continuum limit can be taken, this would resolve the strong CP problem. However, other researches, in which the similar phase diagram was investigated by similar way as Schierholz, indicated that the deconfinement phase boundary may appear due to the statistical errors in their Monte Carlo-based method~\cite{PlefkaMonte, Imachi}. On the other hand, the strong coupling analysis of the phase diagram supported Schierholz's scenario~\cite{Plefka}. Figure \ref{strongcoupling} shows the phase structure of the CP(1) model which we analyze according to their strong coupling method. In the large $\beta$ region, the deconfinement phase appears while the strong coupling analysis is not appropriate for the weak coupling region.

More recently, Azcoiti {\it et al.}~\cite{Azcoiti:2007cg} focused on the fact that the CP(1) model is equivalent to the O(3) model. In the O(3) model, there is a conjecture formed by Haldane which insists that the O(3) model with $\theta =\pi$ is gapless, that is, presents a second-order phase transition~\cite{Haldane:1982rj,Haldane:1983ru}. They investigated whether the same is true or not in the CP(1) model using the Monte Carlo simulation and found that there is a first-order phase transition line for $\beta<0.5$ at $\theta=\pi$, and there is a second-order phase transition line for $0.5<\beta$. The order of the phase transition changes around $\beta=0.5$, but they could not identify the location of the point with a high accuracy.

In general, the Monte Carlo method has a difficulty in analyzing systems including the $\theta$ term due to the sign problem. Therefore it is important to reanalyze the previous results described above by using sign-problem-free methods. The tensor renormalization group (TRG)~\cite{Michael}  method is one of such methods. Thus, we apply this method to the CP($N-1$) model including the $\theta$ term and investigate the phase structure.

In our previous study, a tensor network representation of the CP($N-1$) model including the $\theta$ term was derived and we applyed the TRG method to the CP(1) model without the $\theta$ term~\cite{Kawauchi:2016xng}. In this report, we analyze the phase structure with the $\theta$ term and show the numerical results.

\section{Tensor network representation of the CP($N-1$) model including the $\theta$ term}

In order to apply the TRG method to the CP($N-1$) model, we need a tensor network representation of the partition function. The details of the tensor is given in Ref.~\cite{Kawauchi:2016xng} . Here we only mention the structure of the tensor. 
The dominant term of each elements are as follows.

\begin{figure}[h]
 \centering
  \includegraphics[width=70mm]{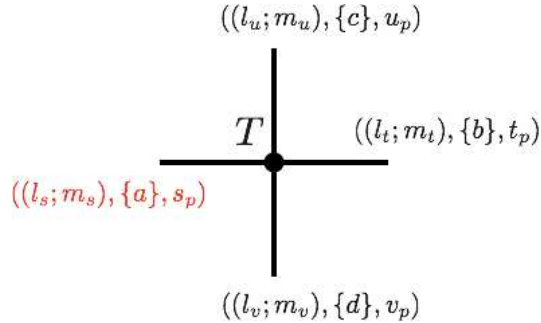}
 \caption{Tensor of the CP($N-1$) model. Each bond has four kinds of indices.}
 \label{cpntensor}
\end{figure}

\begin{align}
\nonumber
T_{stuv}
\equiv&T_{((l_s;m_s),\{a\},s_p)((l_t;m_t),\{b\},t_p)
((l_u;m_u),\{c\},u_p)((l_v;m_v),\{d\},v_p)}\\
&\propto
\sqrt{I_{N-1+l_s+m_s}(2N\beta)}
\times
\frac{2{\rm sin}\frac{\theta +2\pi s_p}{2}}{\theta +2\pi s_p}.
\label{tensor}
\end{align}

\noindent
Focusing on the first bond, there are four kinds of indices, $l_s$, $m_s$, $\{ a\}$, and $s_p$, which are expressed as the red indices in Fig.~\ref{cpntensor}. The indices $l_s$ and $m_s$ are non-negative integers, $\{ a\}$ consists of $\{a_1, a_2, \cdots, a_{l_s+m_s}\}$ where $a_n=1, \cdots, N$, and $s_p$ is an integer. The same is true of the other bonds. 
As the ranges of some of these indices are infinite, one has to truncate the bond dimensions for numerical calculations.
The first factor of Eq.~(\ref{tensor}), $I_{n}(x)$, is the modified Bessel function of the first kind.
One can safely truncate $l_s$ and $m_s$ since the modified Bessel functions of the first kind decrease rapidly as the indices, $l_s$ and $m_s$, increase.
We set the maximum of $l_s + m_s$ to $k$. Table~\ref{dimension} shows how $l_s$ and $m_s$ run and dimension of $\{a\}$ for each set of $(l_s;m_s)$. 

\begin{table}[htbp]
  \caption{Values of $l_s$, $m_s$ and dimension of $\{a\}$.}
  \label{dimension}
  \centering
  \begin{tabular}{ccc}
    \hline
    $l_s$ & $m_s$ & dim. of $\{a\}$\\
    \hline \hline
    0 & 0 & 1\\
    1 & 0 & $N$\\
    0 & 1 & $N$\\
    2 & 0 & $N^2$\\
    1 & 1 & $N^2$\\
    0 & 2 & $N^2$\\
  \end{tabular}\\
  \begin{tabular}{c}
    \vdots \\
  \end{tabular}
\end{table}

\noindent
When the maximum of $l_s + m_s$ is truncated at $k$, the total dimensions of $l_s \bigotimes m_s \bigotimes \{a\}$, which we call $D_\beta$, is

\begin{align}
D_\beta = \frac{1-(2+k)N^{k+1}+(1+k)N^{k+2}}{(1-N)^2}.
\end{align}

\noindent
Especially when $N=2$, $D_\beta=1+k\cdot 2^{k+1}$.
Similarly, one can truncate the dimension of $s_p$ at large $|s_p|$ due to the form of the second factor of Eq.~(\ref{tensor}). We truncate the dimension of $s_p$ to $D_\theta$. We use the truncated tensor for the initial tensor of the TRG method.

\section{Numerical results}

The TRG method makes it possible to calculate the partition function $Z$ approximately once the tensor network representation is obtained. The systematic errors can be controlled by adjusting the bond dimensions of the tensor. This truncation is justified by the hierarchy of the singular values of the tensors. We set the bond dimension to $D_\beta \times D_\theta$. And in this report, we focus on the case $N=2$.

\begin{figure}[h]
 \centering
  \includegraphics[width=110mm]{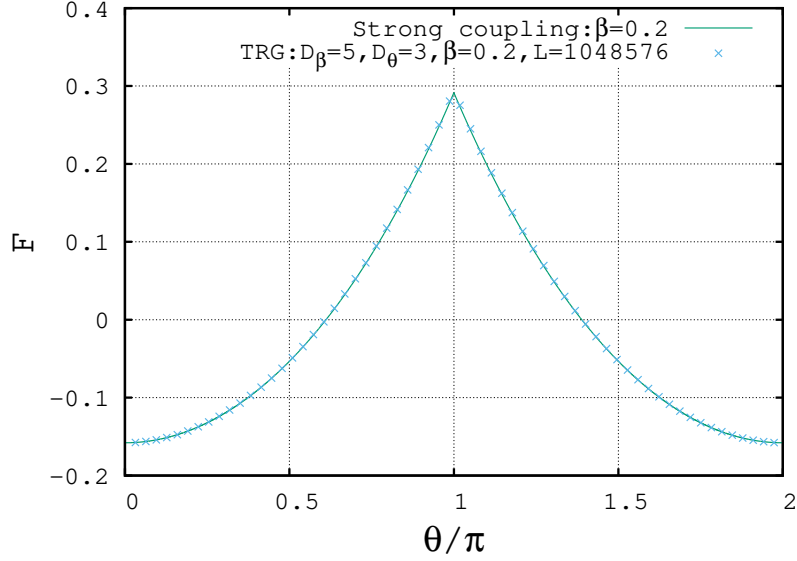}
 \caption{The free energy density $F$ of the CP(1) model at $\beta=0.2$. The solid line indicates the result of the strong coupling analysis and the cross marks indicate that of the TRG method. In the TRG method, the linear lattice size is $L=2^{20}$ and the bond dimension of the tensor is $D_\beta \times D_\theta =5\times 3$.}
 \label{cpn_F_beta02_theta_trg_Linf_Db5_Dt3}
\end{figure}

First, we compare the result of the TRG method with that of the strong coupling analysis. Figure~\ref{cpn_F_beta02_theta_trg_Linf_Db5_Dt3} shows the numerical results of the free energy density $F$ of the CP(1) model at $\beta =0.2$. The strong coupling analysis is expected to work well in small $\beta$ region and these two results are almost consistent at this $\beta$. As the value of $\beta$ grows, however, these two results gradually become different from one another. As can be seen from the results, there seems to be a phase transition at $\theta = \pi$ where the CP symmetry is spontaneously broken~\cite{Seiberg}.

By differentiating twice the free energy density $F$ with respect to $\theta$, one can calculate the topological susceptibility,

\begin{align}
\chi
=
\frac{1}{L^2}\frac{\partial^2}{\partial \theta^2}{\rm log}Z(\theta).
\label{susceptibility}
\end{align}

\begin{figure}[h]
 \centering
  \includegraphics[width=110mm]{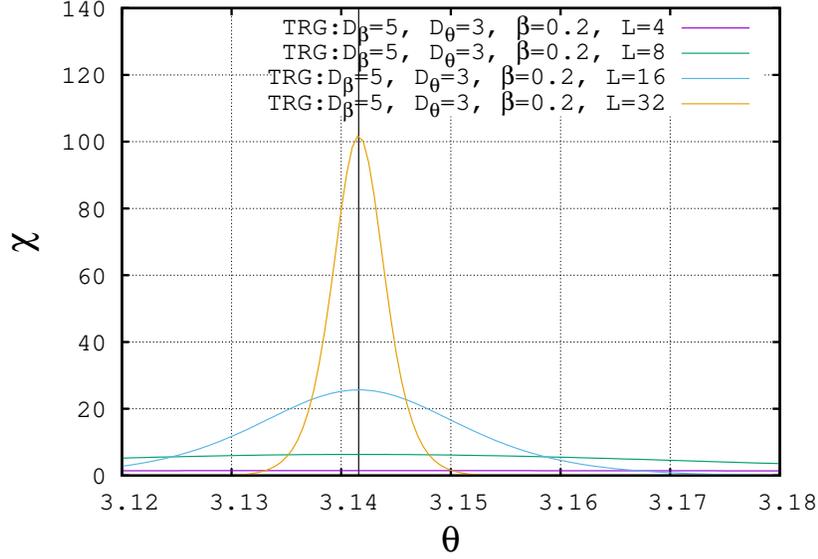}
 \caption{The volume dependence of the topological susceptibility at $\beta=0.2$. The bond dimension of the tensor is $D_\beta \times D_\theta =5\times 3$. The linear lattice size is $L=4$, $8$, $16$ and $32$.}
 \label{cpn_beta02_theta_trg_Db5_Dt3}
\end{figure}

\noindent
We take the derivative with respect to $\theta$ numerically in the TRG method. Figure~\ref{cpn_beta02_theta_trg_Db5_Dt3} shows the volume dependence of the topological susceptibility obtained from the TRG method.

\begin{figure}[h]
 \centering
  \includegraphics[width=107mm]{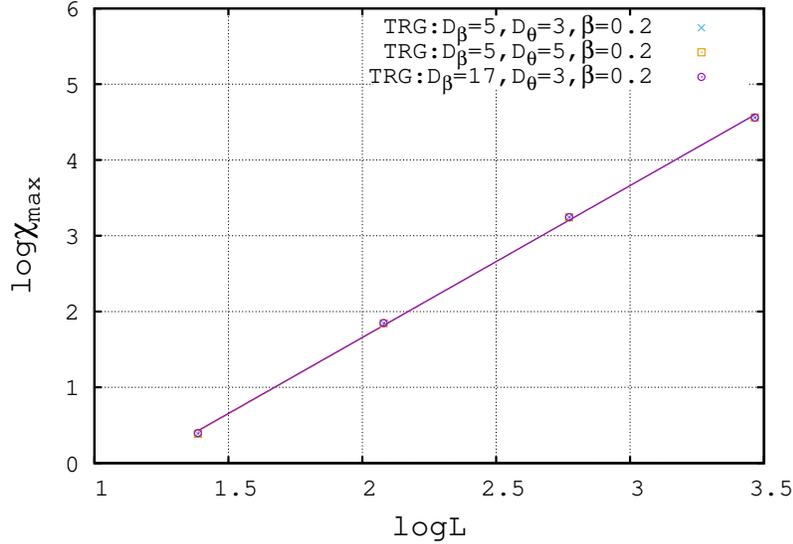}
 \caption{The volume dependence of $\chi_{\rm max}$ at $\beta=0.2$. There is almost no truncation dependence. The slope of the fit line for $D_\beta=17$ and $D_\theta=3$ is $2.003\pm 0.031$ and indicates a first-order phase transition.}
 \label{kaimax_beta02_notitle}
\end{figure}

The order of the phase transition can be verified by the volume dependence of the peak values of the susceptibility, $\chi_{\rm max}$, which is described by
\begin{align}
\chi_{\rm max}\propto L^{\frac{\gamma}{\nu}}.
\end{align}
The exponent $\gamma/\nu$ equals 2 for first-order phase transitions while $\gamma$ and $\nu$ are the conventional critical exponents for second-order phase transitions with $\gamma/\nu <2$.
The result at $\beta=0.2$ is shown in Fig.~\ref{kaimax_beta02_notitle}. In this range of $D_\beta$ and $D_\theta$, the fitting analysis indicates that the phase transition is first order and there is almost no dependence of truncations.

\begin{figure}[h]
 \centering
  \includegraphics[width=110mm]{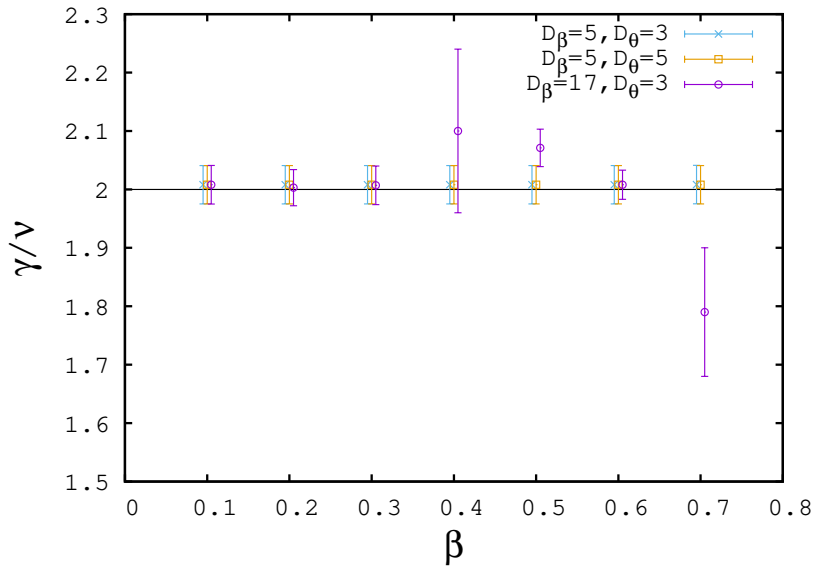}
 \caption{Truncation dependence of the exponent $\gamma / \nu$. For $\beta \leq 0.3$, the truncation dependences are small and it indicates first-order phase transitions. For $0.4\leq \beta$, the truncation dependences are large.}
 \label{kaimax_beta0107_noreverse_error2}
\end{figure}

We also calculate the exponents at different $\beta$s as shown in Fig.~\ref{kaimax_beta0107_noreverse_error2}. This figure shows that the truncation dependences are small in the region $\beta \leq 0.3$, while it tends to be larger in the region $0.4 \leq \beta$. Therefore we have to investigate the latter region with higher accuracy. For that purpose, one can increase the bond dimensions. However it is not reasonable because it is expected that a second-order phase transition line exists from $\beta\sim 0.5$~\cite{Azcoiti:2007cg}. Generally, the TRG method does not work well especially in critical region. For the analysis of critical region, a new tensor network method has already been developed, which is called the tensor network renormalization (TNR)~\cite{Evenbly}. It is worthwhile to apply this method to the critical region.

\section{Summary}

We analyze the phase structure of the CP(1) model including the $\theta$ term by using the TRG method and reconfirm that the order of the phase transition at $\theta=\pi$ is first order until $\beta=0.3$. The truncation dependence of the topological susceptibility arises for $0.4 \leq \beta$. For our future work, we shall try the TNR method in the large $\beta$ region.

\end{document}